\documentclass[10pt,
twocolumn,
amsmath, 
amssymb, 
aps,
prl,
nofootinbib,
raggedbottom,
longbibliography]{revtex4-2}
\usepackage{graphicx} 
\usepackage{preamble}
\usepackage{color}
\usepackage[dvipsnames]{xcolor}
\usepackage{cancel}
\usepackage{tikz-cd}
\usepackage[justification=raggedright,singlelinecheck=false]{caption}
\usepackage{ulem}
\usepackage{subfiles}
\usepackage{tikz-cd}
\usetikzlibrary{decorations.markings}

\newcommand{\avg}[1]{\left\langle#1\right\rangle}

\renewcommand{\Re}{\text{Re\;}}

\begin{document}
 \title{Emergent photons and mechanisms of confinement}
 \author{Jeffrey Giansiracusa}
\author{David Lanners}
\author{Tin Sulejmanpasic}
\affiliation{Department of Mathematical Sciences, Durham University, Durham DH1 3LP, UK \looseness=-1}

\begin{abstract}
We numerically study $\mathbb{Z}_N$ lattice gauge theories in 4D as prototypical models of systems with $\mathbb{Z}_N$ 1-\textit{form symmetry}. For $N \geq 3$, we provide evidence that such systems exhibit not only the expected phases with spontaneously broken/restored symmetry but also a third photon phase. When present, the 1-form symmetry provides a precise notion of confinement, and it is commonly believed that confinement ensues due to the proliferation of extended, string-like objects known as \textit{center vortices}, which carry a $\mathbb{Z}_N$ flux. However, this picture is challenged by the three-phase scenario investigated here. We show that both the confined and the photon phases are associated with the proliferation of center vortices and that the key difference between them lies in whether or not vortex-junctions --- the \textit{monopoles} --- proliferate.
\end{abstract}

\maketitle
\newpage

\section{Introduction}

At center stage of the Standard Model of particle physics, two theories play a distinguished role: the theory of Quantum Electrodynamics (QED), describing light and chemistry, and the theory of Quantum Chromodynamics (QCD), describing the formation of nuclei. While the former is a theory of massless photons that enables us to see far into our universe, the latter is a system with a tiny correlation length of $\sim 1$ fm and is responsible for most of the atomic mass.

Yet both of these theories and regimes are linked to an exotic symmetry called the 1-form symmetry that has been of particular interest in the past decade, mostly starting with the works \cite{Kapustin:2014gua,Gaiotto:2014kfa,Gaiotto:2017yup}\footnote{The 1-form symmetries were known in $\SU(N)$ Yang-Mills theory at least since the late 70s \cite{Polyakov:1978vu}, but until recently were relatively poorly understood.}. On one hand, in QCD, quark confinement is associated with the would-be $\Z_N$ 1-form symmetry --- a symmetry of the pure glue sector --- while the massless photons of QED are seen as Goldstone bosons of a $\U(1)$ 1-form symmetry  \cite{Gaiotto:2014kfa,Hofman:2018lfz}. Therefore, both the massless photon regime and the nuclear confined regime can potentially be understood as phases of matter within the Ginzburg-Landau paradigm for 1-form symmetries (see also \cite{Iqbal:2021rkn}).

In \cite{Nguyen:2024ikq}, it was argued that theories with $\Z_N$ 1-form symmetry in 3+1D have a generic photon phase for any $N$. This phase is seen as an enhancement of the 1-form symmetry, $\Z_N\rightarrow \U(1)$, and consequent spontaneous breaking of it, leading to a massless Goldstone boson: the photon. 

In this work, we will consider $\Z_N$ lattice gauge theory (LGT), a model of classical Ising-like spins on the links of the lattice. These theories have been known to exhibit three phases when $N\ge 5$ \cite{Elitzur:1979uv, Creutz:1979zg}. Here, we show numerically that this is also true for $N=3,4$. 

In parallel, the 3+1D $\Z_N$ LGT is a toy model of pure $\SU(N)$ Yang-Mills theory, having both confined 
and deconfined phases. This model offers a deceptively simple picture of confinement as originating from the proliferation of string-like objects analogous to center vortices of $\SU(N)$ Yang-Mills theory (see \cite{Greensite:2011zz,Greensite:2016pfc} for a review). However, the existence of the photon phase requires this picture to be revised by also considering monopole particles \cite{Stone:1979ug,Nguyen:2024ikq}.


The occurrence of a robust exactly massless Goldstone phase without an exact continuous symmetry may be surprising from the point of view of $0$-form symmetry intuition (see SM for a review). This robustness in 3+1D is what makes the Millennium Problem of whether $\SU(N)$ Yang-Mills theory has a mass gap \cite{Jaffe:2000ne} nontrivial. Indeed, based on the $\Z_N$ $1$-form global symmetry group, we may be erroneously tempted to conclude that a gapless phase is untenable. That a robust photon phase exists invalidates this lore. 

Incidentally, in 2+1D the lore is true, as the photon phase is generically unstable \cite{Nguyen:2024ikq}, resulting in the necessity of a mass gap under all generic symmetry-preserving deformations (see e.g. \cite{Polyakov:1976fu,Unsal:2007jx,Unsal:2008ch,Poppitz:2012sw}).


\section{The $\Z_N$ Lattice Gauge Theory (LGT)}

\begin{figure}[tbp] 
   \centering
   \includegraphics[width=0.45\textwidth]{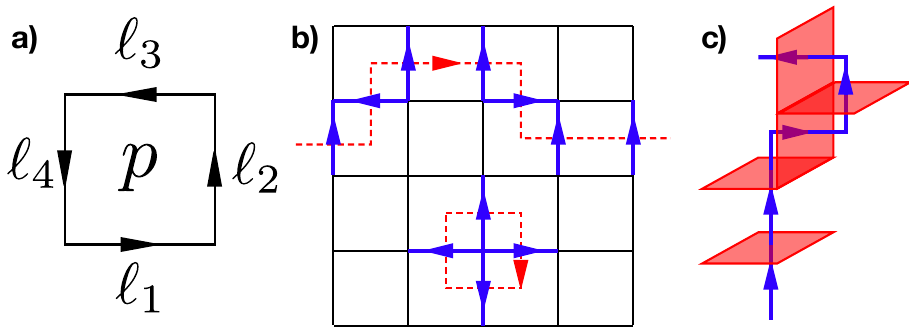} 
   \caption{a) The definition of the plaquette Wilson loop $W_p=U_{\ell_1}U_{\ell_2}U_{\ell_3}U_{\ell_4}$. b) The illustration of the $\Z_N$ 1-form symmetry transformation in 2D. c) The center vortex as an excitation of the plaquette fluxes.}
   \label{fig:definitions}
\end{figure}

The model we consider here is a $\mathbb Z_N$ Lattice Gauge Theory, which is to $\Z_N$ 1-form symmetry what the Ising model is to an ordinary $\Z_2$ symmetry. It is a theory with discrete $N$-state spins $s_\ell=0,1,2,\cdots,N-1$ on lattice links $\ell$. The spins $s_\ell$ naturally form a $\Z_N$ phase $U_\ell=e^{\frac{2\pi i s_\ell}{N}}$. An action that dictates the dynamics of these spins can be constructed by defining $W_p=\prod_{\ell\in\partial p}U_\ell$ --- a plaquette-sized Wilson loop --- where $\partial p$ is the set of boundary links of a plaquette $p$, inheriting the orientation of the plaquette (with the convention $U_\ell=U_{-\ell}^*$). 
The standard Wilson action and the partition function are then given by
\begin{equation}\label{eq:Wilson_action}
S=-\beta\sum_{p}\Re{W_p}\;, \quad
Z=\sum_{\{s\}}e^{-S}\;,
\end{equation}
where the sum is over all spin configurations $\{s_\ell\}$.

A crucial property of this action in general dimension $d$ is that it is invariant under the $\Z_N$ 1-form symmetry \cite{Gaiotto:2014kfa}. To understand this symmetry applied to \eqref{eq:Wilson_action}, consider an oriented $(d-1)$-dimensional hypersurface $H$ constructed out of $(d-1)$-cells of the dual lattice (an example for $d=2$ is given in Fig.~\ref{fig:definitions}b). This will intersect a set of lattice links, which we collect into a set $L$ (marked blue in Fig.~\ref{fig:definitions}b) and orient compatibly relative to $H$. A $\Z_N$ 1-form symmetry is generated by shifting\footnote{We note that by convention $s_{-\ell}=-s_{\ell}\bmod N$. } $s_\ell\rightarrow s_\ell+1 \bmod N$ whenever $\ell\in L$; this is clearly an invariance of the action. 

It is easily checked that for $H$ contractible, the resulting invariance of action is the conventional gauge transformations, while non-contractible choices of $H$ correspond to a global $1$-form symmetry. Thus, the transformations associated with homotopic choices of $H$ are gauge-equivalent, and so the $1$-form symmetry is valued in the co-dimension 1 homology of the space-time manifold with coefficients in $\Z_N$.

The operators charged under the 1-form symmetry are Wilson loops $W[\mathcal C]=\prod_{\ell\in\mathcal C}U_\ell$ where $\mathcal C$ is a contour made out of the links of the lattice. It is easy to see that under the $1$-form symmetry transformation for a hypersurface $H$ the Wilson loop transforms as $W[\mathcal C]\rightarrow W[\mathcal C]\exp\left({i\frac{2\pi I(H,\mathcal C)}{N}}\right)$ where $I(H,\mathcal C)$ is the net intersection number between $H$ and $\mathcal C$. The fact that in the continuum the charged operator is a Wilson loop, which is an integral over the 1-form gauge field $A_\mu$ is what gives the 1-form symmetry its name.

The naive Landau paradigm would then indicate at least two phases: the $\Z_N$ restored (confined) and the spontaneously symmetry broken (SSB) phase, which can be seen respectively in the limits $\beta\rightarrow 0,\infty$. When $0<\beta\ll 1$, the contractible Wilson loops $W[\mathcal C]$ get a leading contribution at order $\beta^{A}$, where $A$ is the area of a minimal surface $S$ whose boundary is $\mathcal C$. By cluster decomposition, this implies that the non-contractible Wilson loops (i.e. Polyakov loops) must have zero expectation value, i.e., the $\Z_N$ 1-form symmetry is restored. In the limit $\beta\gg 1$, the dominant configurations obey $W_p=1$, indicating that non-contractible Wilson loops get a nonzero expectation value.

Further, in this regime, excitations for which $W_p\ne 1$ (see Fig.~\ref{fig:definitions}c) can be thought of as heavy string-like excitations in 3+1D describing a 2D surface on the dual lattice. These string-like objects are analogous to center vortices studied in the QCD community, and this is the name we will adopt. One of the most important features of the center vortex is that whenever it links with a Wilson loop, it is weighted by a $\Z_N$ phase, giving a compelling picture of confinement that can be summarized as follows. As one lowers $\beta$, the center vortex tension goes down, and increasingly larger vortices become favored. At some value of $\beta$, the vortices proliferate, causing dramatic interference with the Wilson loop and eventual area law behavior. This is the essence of the so-called center vortex mechanism of confinement \cite{tHooft:1977nqb,Cornwall:1979hz}.

It has been known since 1979 that for $N\ge 5$, the $\Z_N$ LGT in 4D has a robust phase of emergent photons for intermediate $\beta$ \cite{Elitzur:1979uv,Creutz:1979zg}, making it clear that center vortex proliferation alone cannot possibly explain both the photon and the confined phase. Therefore, a new ingredient is needed here \cite{Nguyen:2024ikq}: the worldline representing a junction of $N$ center vortices --- the monopole-junction.

In fact, a picture of the photon and the confined phase was first briefly touched on in Appendix B of the little-known paper by Stone \cite{Stone:1979ug} and recently in \cite{Nguyen:2024ikq}, which we aim to confirm here. In this framework, both the photon and confined phases involve proliferating center vortices, but differ in the behavior of monopole-junctions: these remain heavy in the photon phase, while they proliferate in the confined phase.

We highlight two important observations:
\begin{itemize}
\item The three phases known to exist for $N\ge 5$ are indeed associated with the (non-)proliferation of the center vortices and monopoles in the way described above and in \cite{Nguyen:2024ikq}
\item The theories for $N = 3,4$ have a deformation which allows for three phases
\end{itemize}
Conspicuously, we will not discuss the theory with $N=2$ in this work because it has no charge-conjugation symmetry and hence no natural choice of monopole-junction. The question of whether theories with (only) $\Z_2$ 1-form symmetry have a photon phase is left for future explorations. 

\section{Numerical simulations and results}\label{sec:model}

\begin{figure}[tbp]
\hspace*{-1cm}%
\scalebox{0.95}{%
 \begin{tikzpicture}[remember picture]
    \node[inner sep=0pt] (image2) at (0,0)
      {\includegraphics[width=0.5\textwidth]{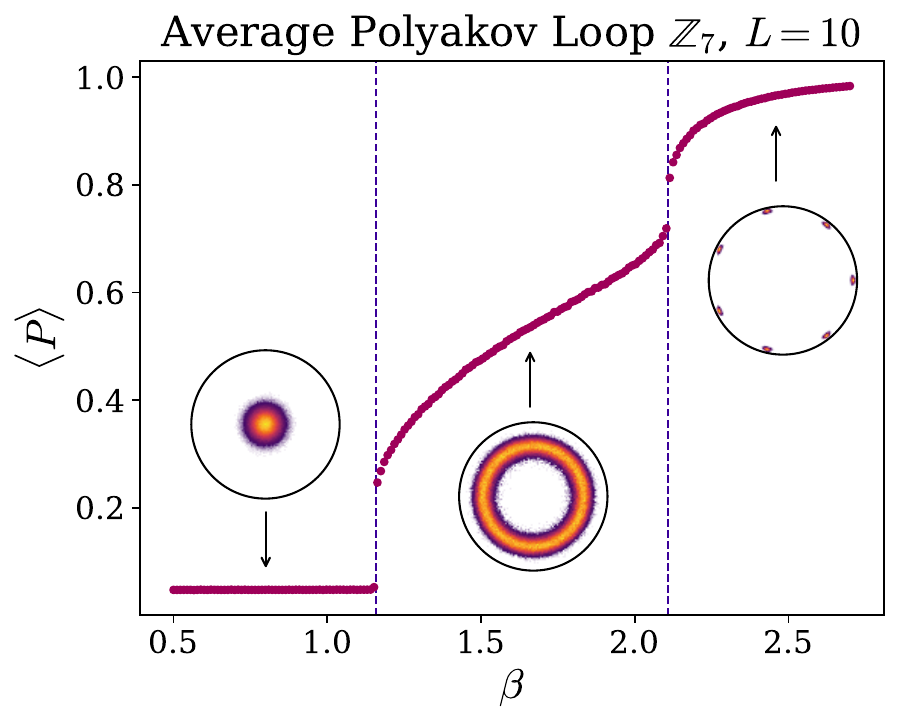}};

    \node[inner sep=0pt] at (-1.9,1.9)
      {\includegraphics[width=0.11\textwidth]{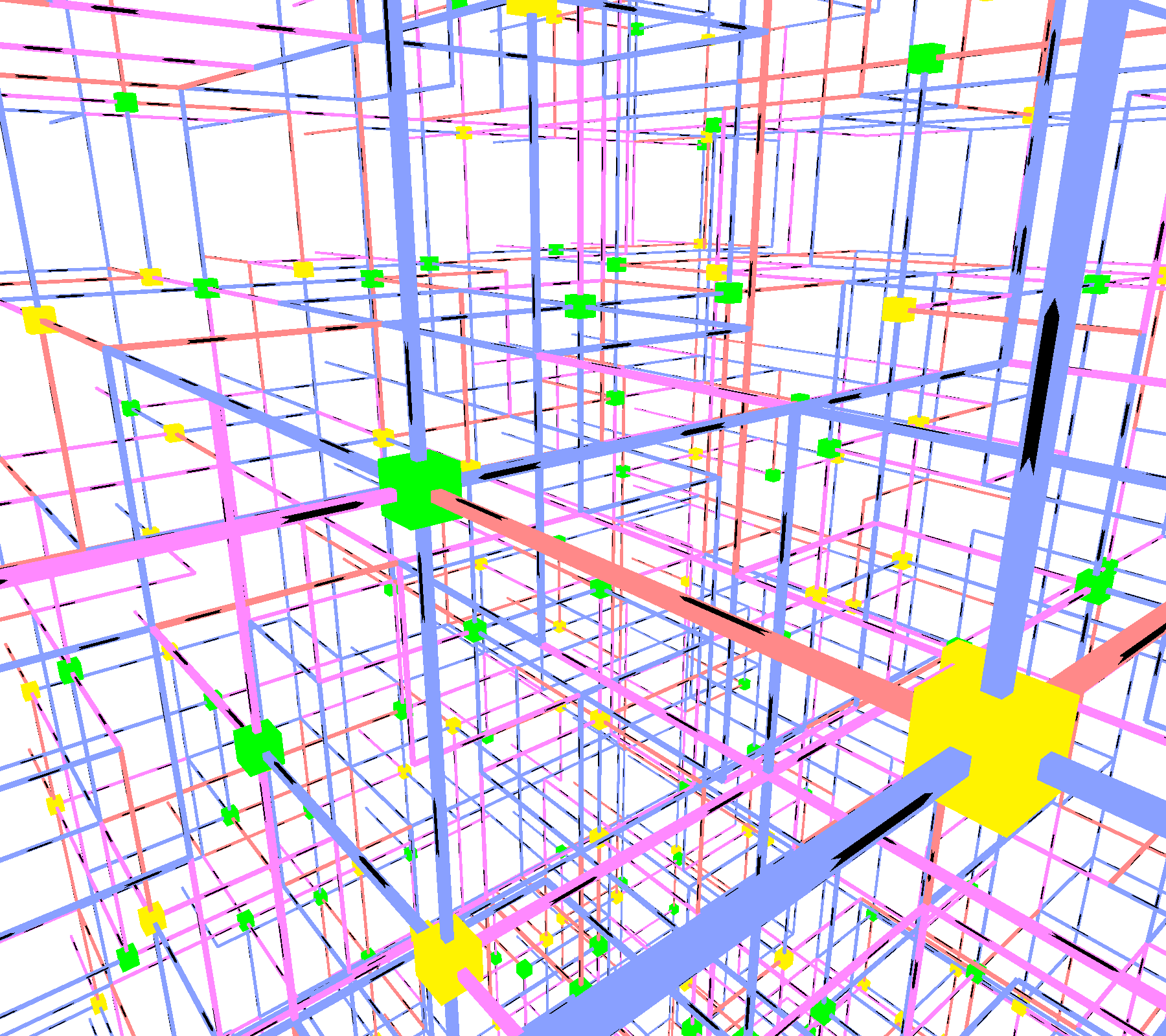}};

    \node[inner sep=0pt] at (0.6,1.9)
      {\includegraphics[width=0.11\textwidth]{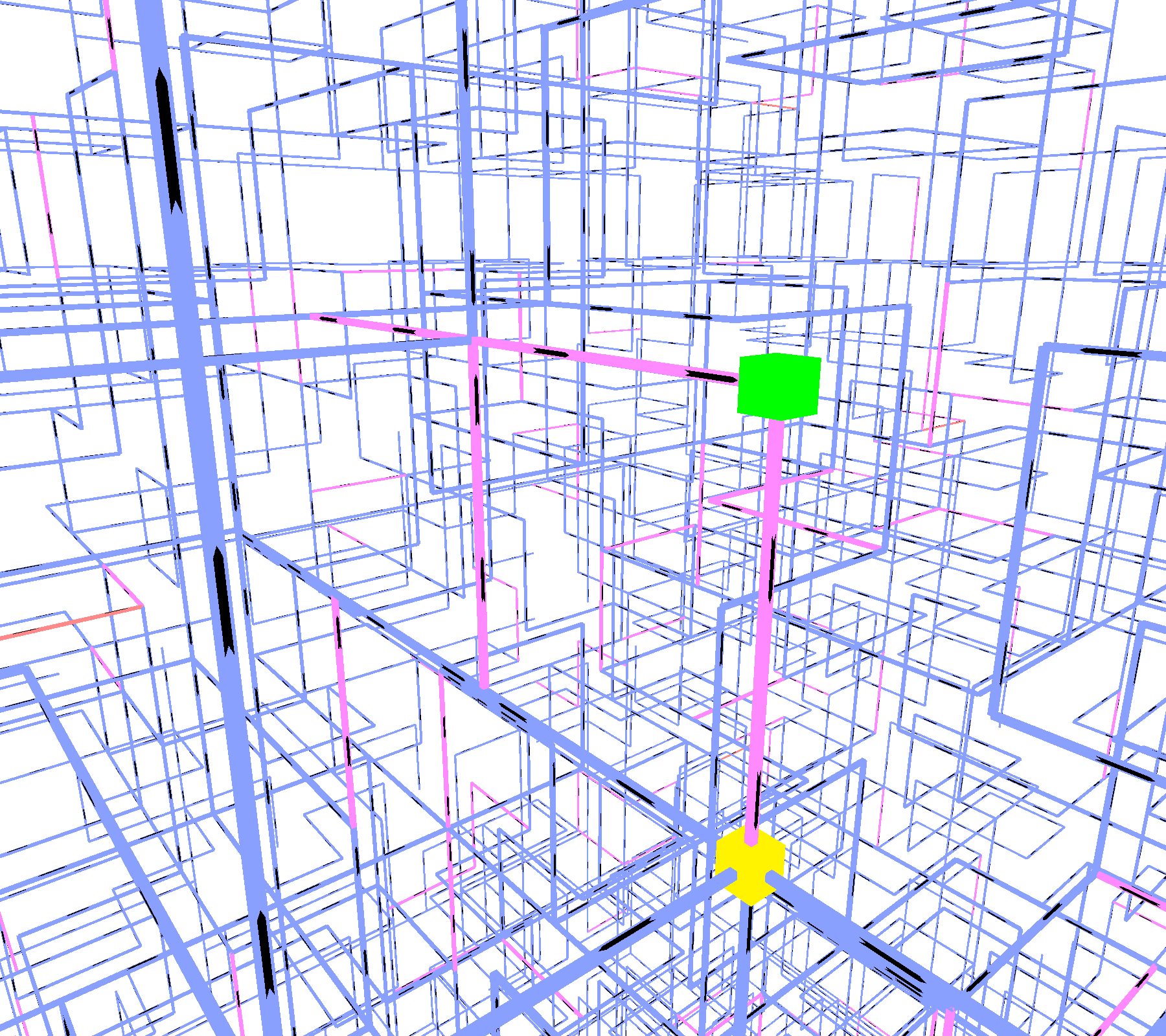}};

    \node[inner sep=0pt] at (3.25,-1.2)
      {\includegraphics[width=0.11\textwidth]{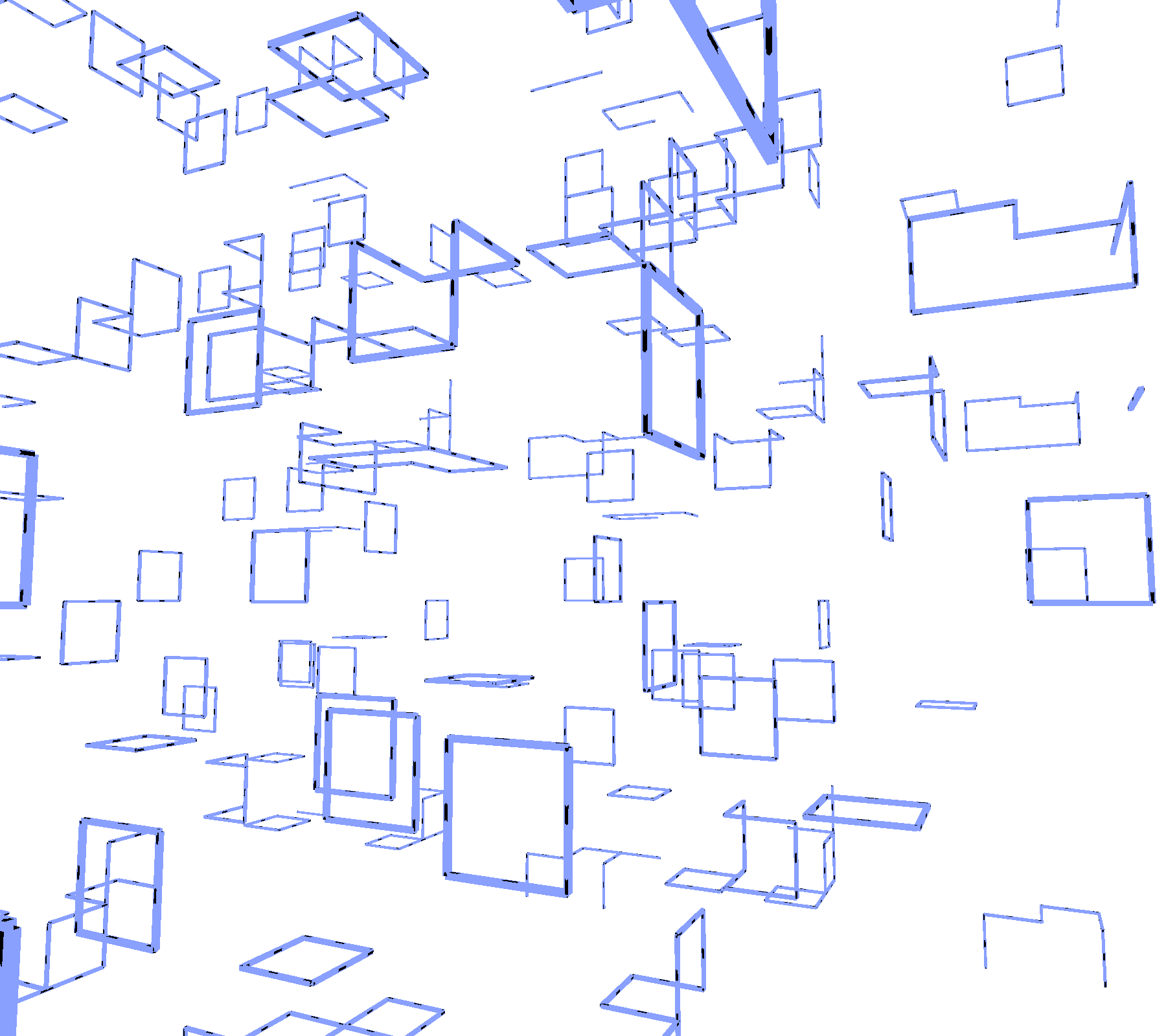}};
 \end{tikzpicture}%
}

\caption{Phase structure of $\Z_7$ Lattice Gauge Theory. Plots show the average Polyakov loop and the corresponding distributions of the smeared Polyakov loop $\bar{P}$. The histograms of $\bar{P}$ indicate confinement (left), emergent U(1) symmetry (middle), and spontaneous $\Z_7$ 1-form symmetry breaking (right). Insets: Typical 3D Monte Carlo configurations showing center vortices (colored by charge ±1, ±2, ±3, arrows indicating the sign) and monopoles/anti-monopoles (green/yellow boxes).}
\label{fig:Z7_phase_diag}
\end{figure}

In this section, we numerically study the two different classes of theories: $N\ge 5$ and $2<N<5$. We choose $N=7$ to represent the $N\ge5$ regime in which three phases are well established \cite{Creutz:1979zg,Elitzur:1979uv}. The goal is to associate these phases with (non-)proliferating center vortices and monopole-junctions. To this end, we begin by formally defining these objects on the lattice.

The center vortex is easy to define by considering a flux through a plaquette $p$ given by $W_p=\exp\left(i\frac{2\pi f_p}{N}\right)$, where $f_p$ is an integer well-defined modulo $N$, for which we choose the unique representative in the range:
\begin{equation}
-\frac{N}{2} < f_p\le \frac{N}{2}\;.
\end{equation}
We have $f_p=(ds)_p+Nm_p$, where $m_p$ is some integer defined on plaquettes (a 2-cochain). Further, since $d^2=0$ (see e.g. \cite{wallace2007algebraic,Sulejmanpasic:2019ytl}), we have that $(df)_c=\sum_{p\in \partial c}f_p\equiv 0\bmod N$, with $c$ denoting a cube of the lattice. 

In Fig.~\ref{fig:definitions}c), we give an example of a center vortex configuration by shading the plaquettes for which $f_p\equiv 1\bmod N$. If $(df)_c=0$ were to hold, then the center vortex would form closed string worldsheet configurations on the dual lattice. However, because $(df)_c\equiv 0\mod N$, there can be configurations for which $(df)_c=N(dm)_c$, so $(dm)_c\ne0$ (a loop on the dual lattice) is our definition of a monopole-junction -- a worldline where $N$ center-vortex worldsheets meet.

Consider now the $\Z_7$ LGT with the usual Wilson action \eqref{eq:Wilson_action}.
To probe this model, we ran Monte Carlo simulations for 200 values of $\beta\in [0.5,2.7]$, and performed $2\cdot10^4$ measurements each, with 20 decorrelation sweeps between measurements (see SM for more details on the simulations). Furthermore, we chose $L=10$ and estimated the error via bootstrap. We present the results in Fig.~\ref{fig:Z7_phase_diag}. Note that the errors are too small to be seen. 

This theory has three phases, which we can distinguish by the Polyakov loop (i.e. the Wilson loop winding along one of the cycles of the torus). Fig.~\ref{fig:Z7_phase_diag} shows the Polyakov loop expectation value, which shows a clear 1st order jump at two values of $\beta_1\approx 1.157\pm 0.006$ and $\beta_2\approx 2.108\pm 0.006$. The confined phase appears for small values where the Polyakov loop expectation value is (nearly) zero, but the middle (the photon) and the large $\beta$ (the SSB) phases both have a finite Polyakov loop VEV. To distinguish the two, we measure the following. Let $x=(t,x_1,x_2,x_3)$ parametrize the 4D position on a torus (i.e. $t\sim t+L$ and $x_i\sim x_i+L$). Let $\mathcal L_{\bm x}$ be the sequence of all links pointing in the temporal direction at the spatial point $\bm x=(x_1,x_2,x_3)$. Now, define the temporal Polyakov loop to be $P_{\bm x}=\prod_{\ell\in \mathcal L_{\bm x}}U_\ell$. To each configuration we assign a spatially smeared Polyakov loop $\bar P=\frac{1}{L^3}\sum_{\bm x}P_{\bm x}$, and plot its distribution across the Monte Carlo measurements for the typical representatives of the three phases in Fig.~\ref{fig:Z7_phase_diag}, as well as 3D slices of the corresponding configurations. 

In the SSB phase $\beta>\beta_2$, the smeared Polyakov loop shows a $\Z_7$ symmetric distribution reflecting a spontaneously broken $\Z_7$ 1-form symmetry. In this phase, the center vortices correspond to short loops. As $\beta$ is lowered to $\beta_1\lesssim\beta\lesssim \beta_2$, the system transitions into the photon phase, the center vortices proliferate, and $\bar P$ develops a circular distribution, indicating the emergence of the $\U(1)$ 1-form symmetry. Notably, the monopoles -- depicted by green and yellow cubes in the figure -- are uncommon, indicating that monopole worldlines are short and non-proliferating. As $\beta$ is reduced further to $0<\beta<\beta_1$, the confined phase sets in, whose hallmark is a distribution of $\bar P$ unsurprisingly sharply peaked around zero, and a proliferation of both center vortices as well as monopoles, consistent with the proposal of \cite{Nguyen:2024ikq}. 

\begin{figure}[!tbp] 
   \centering
   \hspace{-8mm}
   \includegraphics[width=0.48\textwidth]{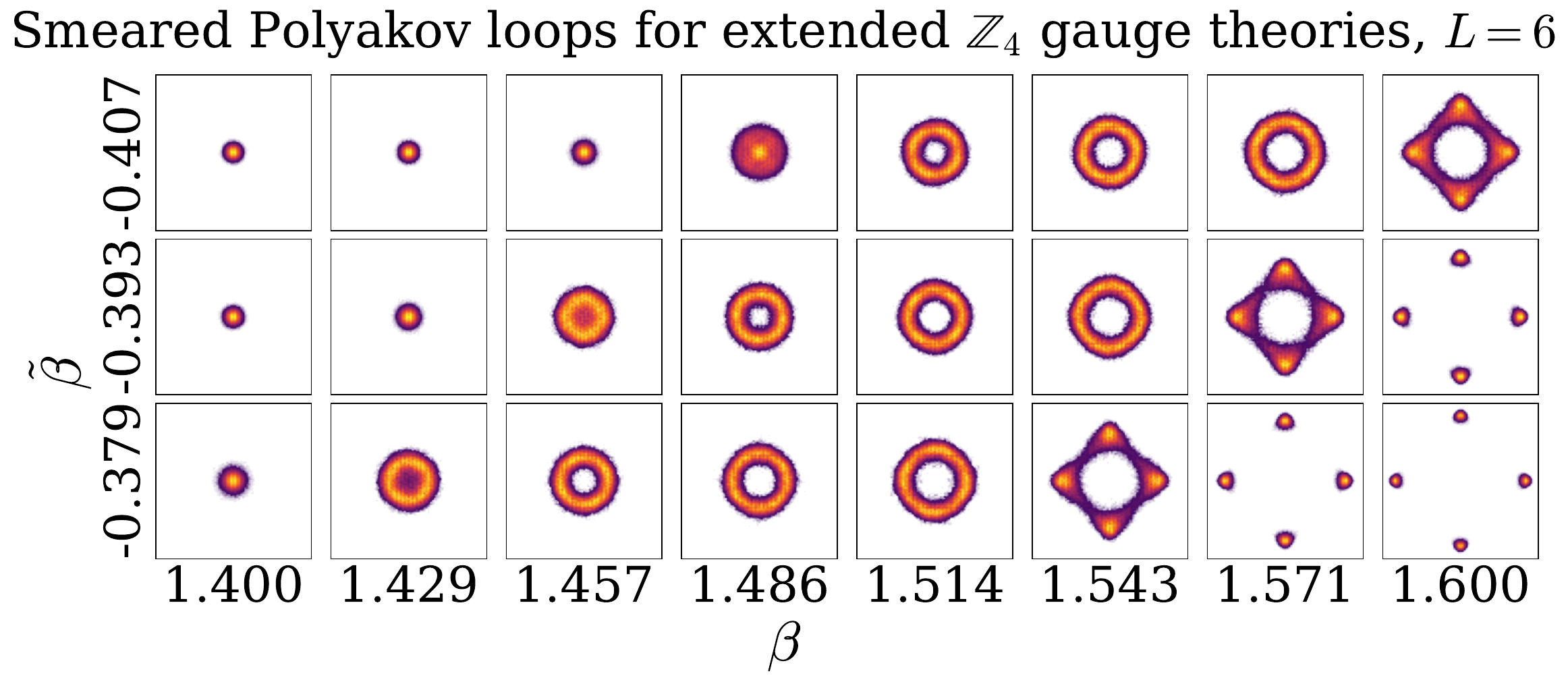} 
   \caption{Histogram of the smeared Polyakov loop as a function of $\beta$ and $\tilde\beta$ (see \eqref{eq:S-Z4}) of the $\Z_4$ lattice gauge theory.}
   \label{fig:Z4}
\end{figure}

This behavior is expected for any $N\ge 5$. However, in \cite{Alcaraz:1982xe,Creutz:1982dn,Fukugita:1982kk}, a $\Z_4$ theory was studied, with the following modified action
\begin{equation}\label{eq:S-Z4}
S=-\beta \sum_p\Re W_p-\tilde \beta \sum_p \Re W_p^2\;.
\end{equation}
They argued that for a particular choice of $\beta$ and $\tilde\beta$, a massless photon phase emerges. Indeed in Fig.~\ref{fig:Z4} we present the Monte Carlo time series histogram of the smeared Polyakov loops in Fig.~\ref{fig:Z4}. As can be observed, we see the same pattern of circle formation indicating a photon phase in between the confined and the SSB phase, confirming the photon phase.

Now we turn to the $\Z_3$ model. In this model with the standard Wilson action \eqref{eq:Wilson_action}, there are only two phases. The picture of this model is that as $\beta$ decreases from $\beta\gg 1$ to $\beta\ll 1$, both center vortices and monopole-junctions become abundant simultaneously. To open a photon phase, we therefore add a monopole suppression term to the action:
\begin{equation}
S=-\beta \sum_p\Re W_p+\mu\sum_c (dm)_c^2\;.
\end{equation}

\begin{figure}[tbp] 
\begin{center}
 \begin{tikzpicture}[remember picture]  
    \node[inner sep=0pt] (image2) at (0,0)
      {\includegraphics[width=0.5\textwidth]{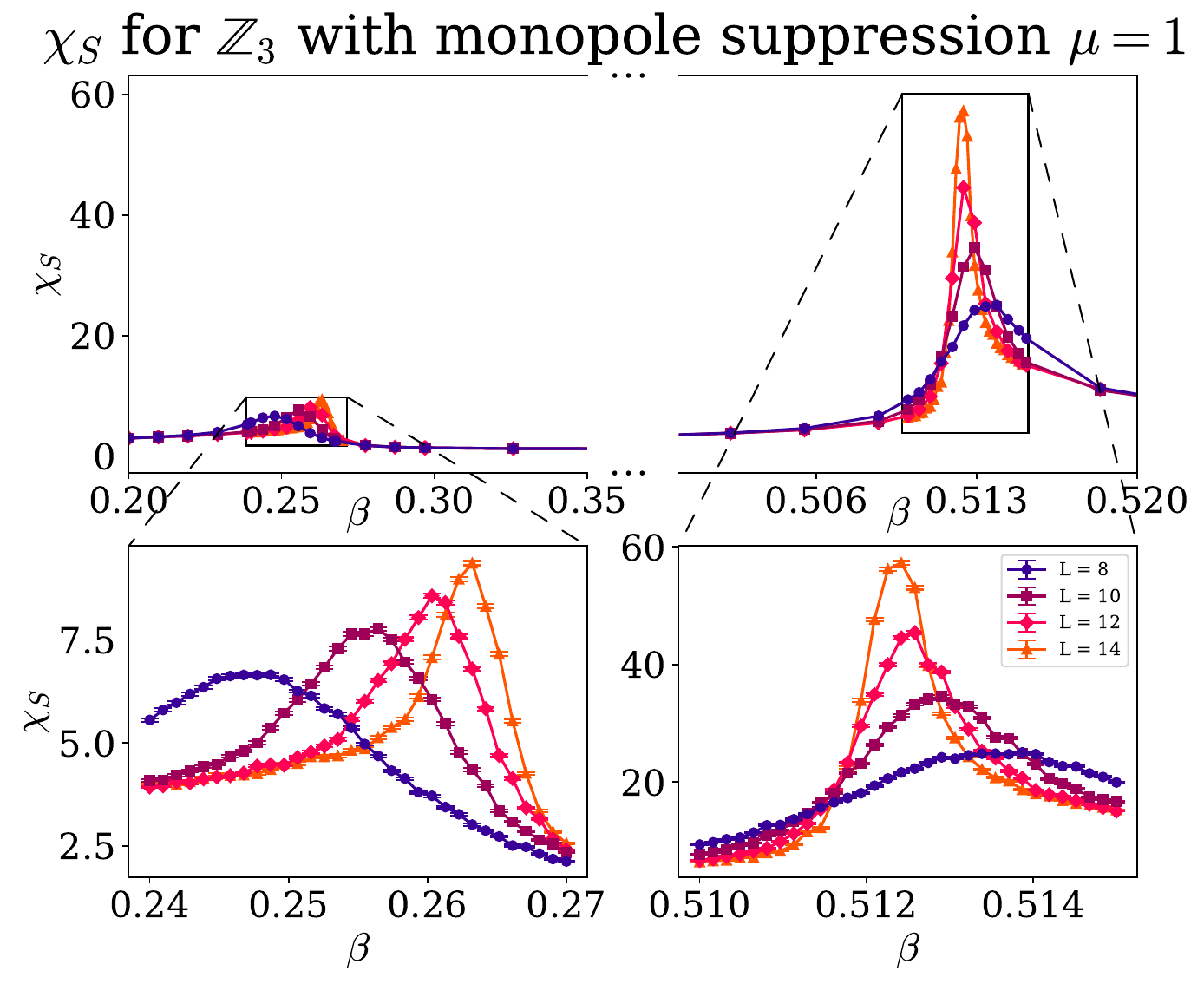}};

      \node[inner sep=0pt] (image2) at (-1.6,1.92)
      {\includegraphics[width=0.21\textwidth]{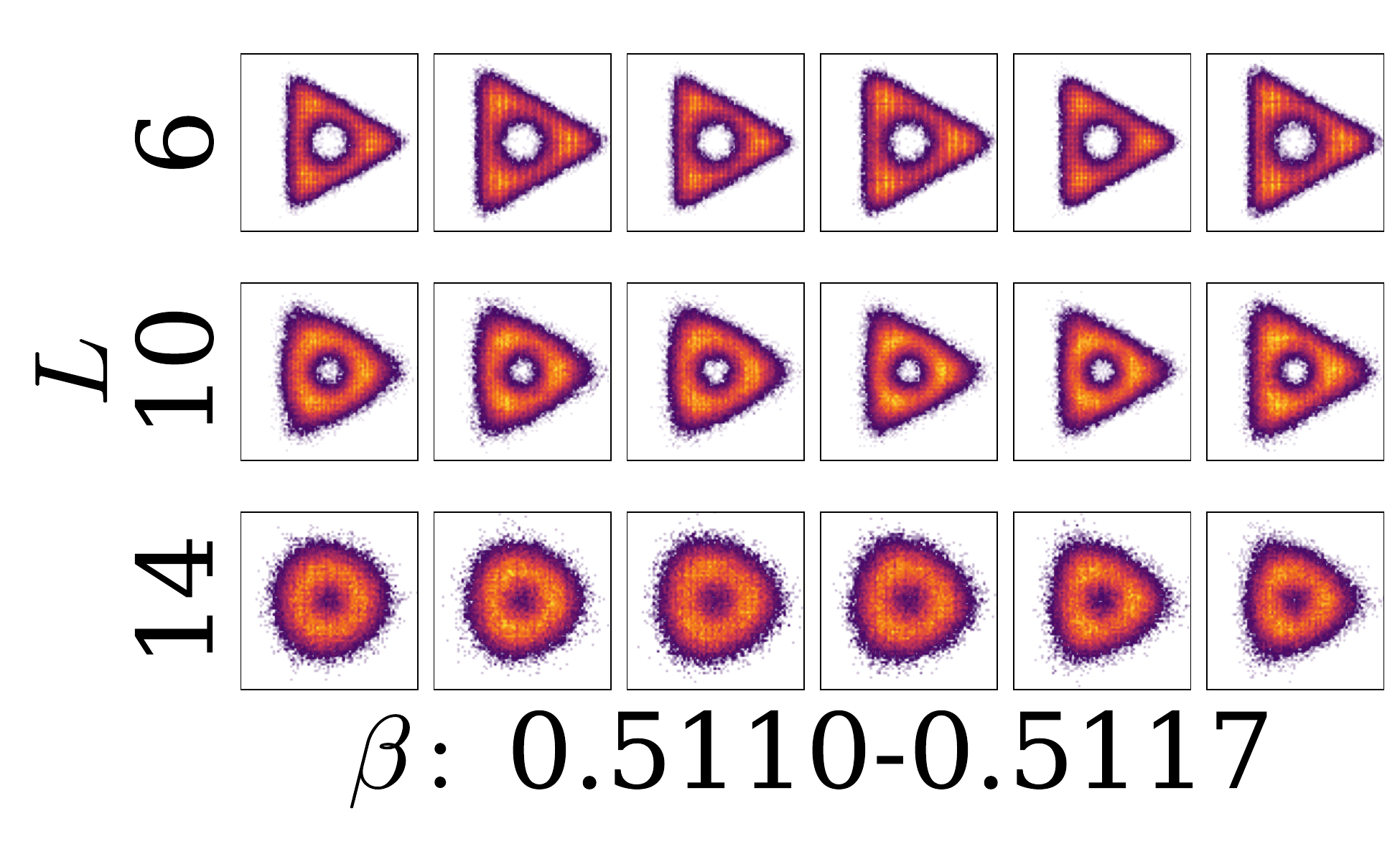}};
  \end{tikzpicture}
  \captionof{figure}{{Susceptibility $\chi_S$ versus $\beta$ for the $\Z_3$ model with monopole suppression ($\mu$=1) for lattices $L=8,10,12$ and $14$. Two prominent peaks indicate phase transitions. Bottom panels show zooms of the low-$\beta$ (confined $\to$ photon) and high-$\beta$ (photon $\to$ SSB) transition regions. Inset: smeared Polyakov loop histograms near the high-$\beta$ transition for $L=6, 10$ and $14$.}}\label{fig:S_susceptiblity}
\end{center}
\end{figure}

To identify phase transitions, we look at the action susceptibility (``heat capacity'') defined as
\begin{equation}
\chi_S=\frac{1}{L^4}\left(\avg{S^2}-\avg{S}^2\right).
\end{equation}
In Fig.~\ref{fig:S_susceptiblity} we plot numerical Monte Carlo results for $\chi_S$ as a function of $\beta$ at $\mu=1$.
We ran Monte Carlo simulations for $L=8,10,12,$ and $14$ for various values of $\beta$ (for simulation details see SM). The results for $\chi_S$ can be seen in Fig.~\ref{fig:S_susceptiblity}. As is clear, the susceptibility shows two peaks, indicating a three-phase scenario, with phase transitions at $\beta_1\approx 0.265$ and $\beta_2\approx 0.512$. The inset shows the smeared Polyakov loop $\bar P$ for values of $\beta$ close to the second transition, showing the circles forming in the photon phase.

Finally, we also consider the connected plaquette-plaquette correlators (PPC) defined as (no sum over $\mu,\nu$)
\begin{equation}
C^\pm_{x-y;\mu\nu}=\avg{W_{p_{x,\mu\nu}}^{\pm 1}W_{p_{y,\mu\nu}}}-\big\langle W_{p_{x,\mu\nu}}^{\pm 1}\big\rangle \big\langle W_{p_{y,\mu\nu}}\big\rangle\;,
\end{equation}
where $W_{p}^{-1}=1/W_p$, and where $p_{x,\mu\nu}$ indicates a plaquette with a corner at lattice site $x$, expanding in directions $\mu,\nu=0,1,2,3$, and averages are over ensemble and all spatial translations.
At long distances in the photon phase, the theory flows to a pure $\U(1)$ gauge theory with the action $S=\frac{1}{4}\int d^4x\; F_{\mu\nu}F^{\mu\nu}$, which describes free photons. All gauge invariant local operators of the $\U(1)$ gauge theory must be built from the field-strength operator $F_{\mu\nu}$, so we expect $C_{x-y;\mu\nu}\propto \avg{F_{\mu\nu}(x)F_{\mu\nu}(y)}$ for large $|x-y|$. Further, we have that
\begin{equation}
    \avg{F_{\mu\nu}(x)F_{\rho\sigma}(y)}\propto\frac{D_{\mu\nu;\rho\sigma}}{|x-y|^4}\;,
\end{equation}
where $D_{\mu\nu;\rho\sigma}$ is independent of the modulus $|x-y|$ and is easily computed\footnote{One way is to note that $\avg{A_\mu(x)A_\nu(y)}=\delta_{\mu\nu}/|x-y|^2$ in the Feynman gauge, and then take derivatives.}, but is not relevant for our discussion.  Let us define a combination\footnote{The combination is chosen because, by reflection positivity, it is a sum of positive definite parts.}
\begin{equation}\label{eq:asympt}
    C(n)=\!\!\sum_{\substack{\rho,\mu,\nu\\\mu>\nu\ne \rho}}C^-_{n\hat \rho,\mu\nu}\!+\!\sum_{\substack{\rho,\mu\\\rho\ne \mu}}C^+_{n\hat \rho,\rho \mu}\approx \underbrace{\frac{K}{n^4}\!+\!\frac{K}{(L-n)^4}}_{C_{\text{asym}}(n)}\;,
\end{equation}
where $\hat{\rho}$ is a lattice unit vector in the direction $\rho$, and $K$ is a constant, and where we took into account the corrections due to the finiteness of the periodic lattice.


Since in the ratio $C(n)/C(n+1)$ the constant $K$ cancels, all correlators in a Coulomb phase should asymptote to a universal function. In Fig.~\ref{fig:power-law} we show numerical results of this ratio for $\Z_{3},\Z_4$ and $\Z_7$ in the Coulomb phase and find excellent asymptotic agreement, confirming the photon phase of the model. In the SM, we also show independent correlators $C^{\pm}(n)$ and find that they perfectly fit the expansion into powers of $1/n^2$. 
\begin{center}
 \begin{tikzpicture}[remember picture]  
    \node[inner sep=0pt] (image2) at (0,0)
      {\includegraphics[width=0.5\textwidth]{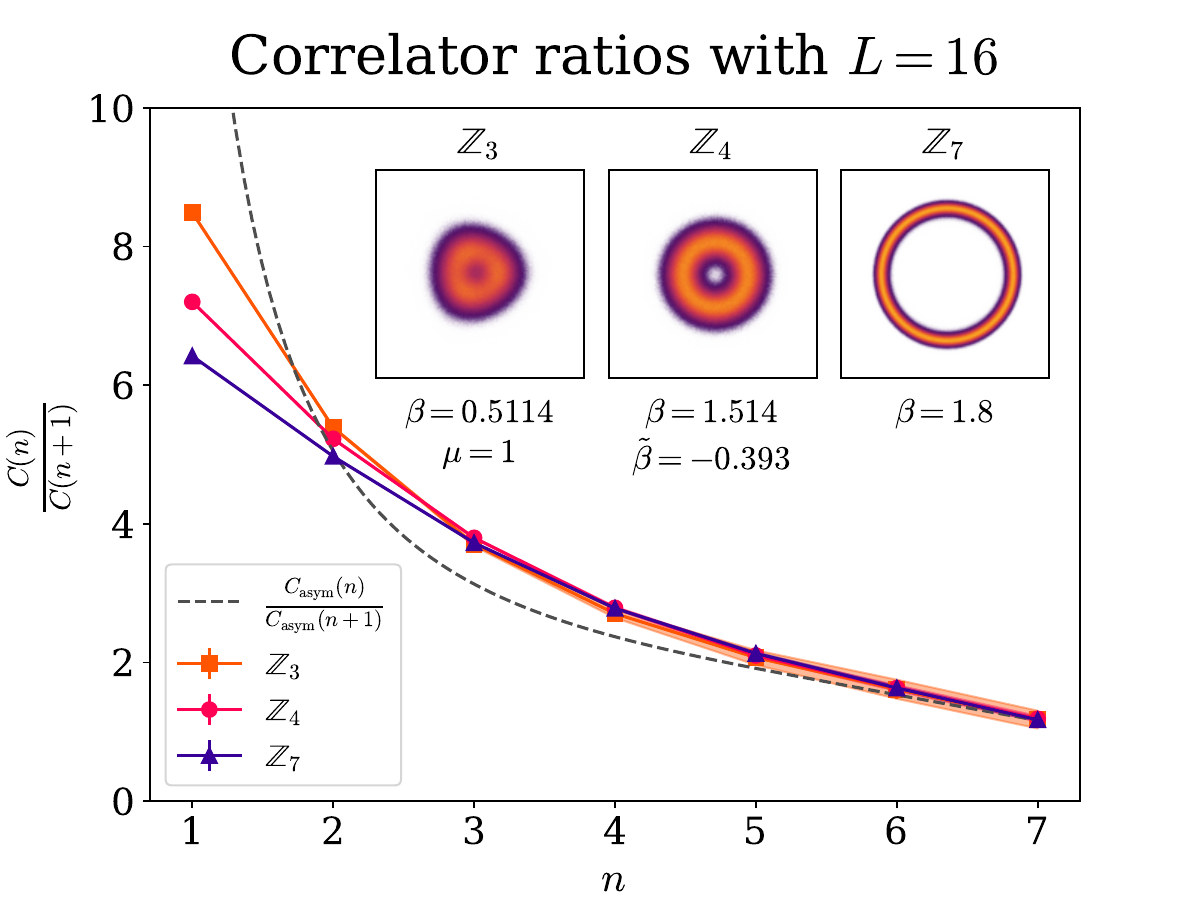}};
  \end{tikzpicture}
 \captionof{figure}{The correlator ratio $C(n)/C(n+1)$ in the photon phases of $\Z_3, \Z_4$, and $\Z_7$ on an $L=16$ lattice compared to the long-distance expectation (see \eqref{eq:asympt}). Insets: Corresponding smeared Polyakov loops.} \label{fig:power-law}

\end{center}

\section{Conclusions}

We numerically showed that $\Z_N$ lattice gauge theories have three phases for all $N\ge 3$, supplementing the spontaneously broken and restored $\Z_N$ 1-form symmetry with a phase of emergent photons where the symmetry enhances to $\U(1)$. 

In addition, we demonstrated a direct link between the photon phase and the proliferation of string-like center vortices, while the confined phase was linked to proliferation of both vortices and vortex junctions (monopoles) as proposed in \cite{Nguyen:2024ikq}. This suggests that center vortices are not directly responsible for confinement, as commonly believed for $\SU(N)$ gauge theory.\footnote{Defining monopoles and center vortices requires gauge fixing in $\SU(N)$ gauge theories (see e.g. \cite{Chernodub:1997ay,Greensite:2011zz,Greensite:2016pfc}), although there are some efforts to identify such objects without it \cite{Sale:2022qfn,Sale:2022qfn}. Further, a novel proposal of how to discretize $\SU(N)$ gauge theories \cite{Zhang:2024cjb,Chen:2024ddr}, may be related to a gauge invariant definition of vortices and monopoles \cite{private_comm}.}

Our work indicates that the Landau paradigm for $\Z_N$ 1-form symmetries must include a massless phase when $N\ge 3$, distinguishing it from its 0-form counterpart.

The case of $\Z_2$ lattice gauge theory is more subtle as it does not have a global charge conjugation symmetry $C$.\footnote{TS would like to thank M.~Nguyen and E.~Richards for discussions on this point.} While it was proposed in \cite{Nguyen:2024ikq} to have a photon phase when deformed, such deformations would introduce the $C$-symmetry. A similar issue occurs in dihedral group ($D_N$) lattice gauge theories, which have no $C$ symmetry and only at most a $\Z_2$ 1-form symmetry. Yet for large enough $N$, $D_N$ LGT should be indistinguishable from $O(2)$ LGT, which certainly has a photon phase, albeit with a non-invertible emergent 1-form symmetry\footnote{The lattice simulations of $D_N$ seem to indicate three phases for $N\gtrsim 6$ \cite{Alam:2021uuq}.}. The Landau paradigm for $\Z_2$ symmetry therefore likely involves a massless phase as suggested in \cite{Nguyen:2024ikq}, although the details are more subtle and are left for future explorations.

\section*{Acknowledgement}
We would like to thank Mathew Bullimore, Xavier Crean, Christof Gattringer, Tyler Helmuth, Nabil Iqbal, Zohar Komargodski, Biagio Lucini, Mendel Nguyen, Emily Richards, Shu-Heng Shao and Yuya Tanizaki for the discussions and comments. This work has made use of the Hamilton HPC Service of Durham University. TS is supported by the Royal Society University Research Fellowship and, in part, by the STFC consolidated grant number ST/T000708/1. JG is supported by EPSRC grant EP/Y028872/1.

\bibliography{refs.bib}

\newpage
\begin{appendix}

\begin{center}
\vspace{2cm}
    {\bf \Large Supplementary Material}
\end{center}

\section{Details of simulations}

All measurements were generated on a 4-torus of lattice size $L^4$, interpreted as Euclidean space-time, using Monte Carlo simulations with sequential sweeps of single-link Metropolis updates, where each link is updated once per sweep. A trial link value, chosen uniformly from $\mathbb{Z}_N$, is accepted with probability $\min\{1,e^{-\Delta S}\}$ based on the local action change $\Delta S$; otherwise, the original value is retained. Table \ref{tab:sim_details} shows the details of lattice simulations presented in the Figures of the main text.

\begin{table}[h]
\centering
\begin{tabular}{|l|c|c|c|c|}
\hline
 & \textbf{Figure 2} & \textbf{Figure 3} & \textbf{Figure 4} & \textbf{Figure 5} \\
\hline
Theory  & $\mathbb{Z}_7$ & $\mathbb{Z}_4$ & $\mathbb{Z}_3$ & $\mathbb{Z}_3$, $\mathbb{Z}_4$, $\mathbb{Z}_7$ \\
\hline
Lattice sizes   & $10$ & $6$ & $8, 10, 12, 14$ & $16$ \\
\hline
\# measurements & $2 \times 10^{4}$ & $10^{4}$ & $4 \times 10^{4}$ & $1.28\times 10^6$ \\
\hline
Decor. sweeps   & 20 & 1 & 100 & 500 \\
\hline

\end{tabular}
\caption{Simulation details.}\label{tab:sim_details}
\end{table}

\section{Binned Jackknife Error Estimation}
Consecutive Monte Carlo samples are generally correlated, so straightforward error estimates tend to underestimate uncertainties. To address this, we divide the data into $N_b$ contiguous bins of size $B$ \cite{Janke}, chosen large enough such that correlations between the bins can be neglected. Jackknife samples are then formed by omitting one bin at a time and averaging over the remaining $N_b-1$ bins, with the variance of the resulting estimates rescaled by $(N_b-1)/N_b$ to compensate for the fact that each jackknife sample is based on $N_b-1$ bins rather than the full set, which would otherwise bias the variance downward. Let $\mathcal{O}$ be any observable and denote by $\overline{\mathcal{O}}_{B,1}, \dots , \overline{\mathcal{O}}_{B,N_b}$ the bin averages. The jackknife estimate from leaving out bin $i$ and the jackknife mean are given by $$\mathcal{O}_i = \frac{1}{N_b-1} \sum_{\substack{j=1 \\ j\neq i}}^{N_b} \overline{\mathcal{O}}_{N,j} \ \ \text{ and } \ \ \overline{\mathcal{O}} = \frac{1}{N_b} \sum_{i=1}^{N_b} \mathcal{O}_i.$$ The statistical error can then be obtained by:
$$\sigma_{\mathcal{O}}^2 = \frac{N_b -1}{N_b} \sum_{i=1}^{N_b} (\mathcal{O}_i - \Bar{\mathcal{O}})^2,  $$ In practice, the bin size is increased successively, $B = 2, 5, 10,20,50,100, 200,500,\dots$, and once the error saturates the corresponding binned averages can be treated as effectively independent. 

\begin{figure}[tbp] 
\begin{center}
\begin{tikzpicture}[remember picture]  
\node[inner sep=0pt] (image2) at (0,0)
{\includegraphics[width=0.45\textwidth]{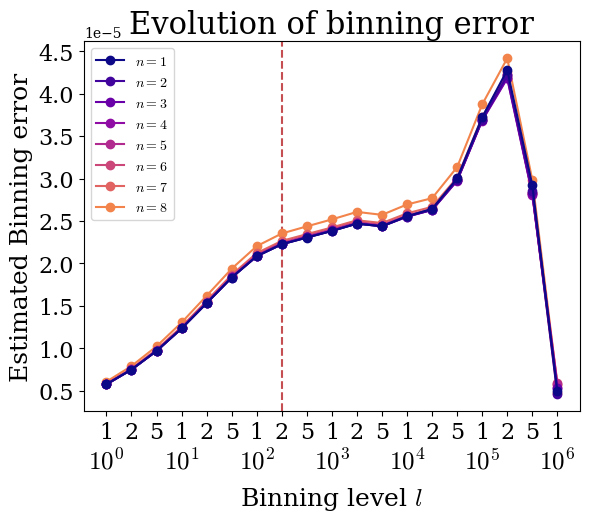}};
\end{tikzpicture}
\captionof{figure}{Saturating binning error for longitudinal correlators. The dashed line marks the onset of the plateau, motivating a choice of 500 decorrelation sweeps.}\label{fig:error-evolution}
\end{center}
\end{figure}

This saturation point also provides a useful heuristic for approximating how many sweeps should be skipped between recorded configurations in order to obtain uncorrelated samples. Effectively, we run the simulation with a given observable in mind, monitor the jackknife error as the bin size increases, and use the onset of a plateau to assess autocorrelation. If the plateau only emerges at large bin sizes, this indicates that configurations remain correlated over many sweeps, and the number of skipped sweeps should be increased accordingly. By doing so, we avoid recording correlated configurations, thereby reducing memory and computational costs while maintaining statistical accuracy. The evolution of the binning error for the longitudinal correlators can be seen in Figure~\ref{fig:error-evolution}.

\section{Smeared Polyakov Loops}
Recall that the temporal Polyakov loop was defined to be $P_{\boldsymbol{x}}= \prod_{l\in \mathcal{L}_{\bold x}} U_l \in \Z_N$. Let $\varphi: \mathbb{Z}_N \to \mathbb{R}^2$ be the map that identifies $\Z_N$ with the complex roots of unity, and then regards them as points in $\mathbb{R}^2$. The smeared Polyakov loop for a given configuration is then given by taking the spatial average and interpreting it in $\mathbb{R}^2$:
$$ \overline{P} = \frac{1}{L^3} \sum_{\bold x} \varphi (P_{\bold x}) \in \mathbb{R}^2.$$One immediate consequence of this definition is that the smeared Polyakov loop is contained in the $N$-gon spanned by the $N$-th roots of unity. 

\begin{center}
\begin{tikzpicture}[scale=1.5]

\draw[line width=0.5pt, red, dashed] (1, 0) -- (0.1, 0.2);
\draw[line width=0.5pt, red, dashed] (0.309, 0.951) -- (0.1, 0.2);
\draw[line width=0.5pt, red, dashed] (-0.809, 0.587) -- (0.1, 0.2);
\draw[line width=0.5pt, red, dashed] (-0.809, -0.587) -- (0.1, 0.2);
\draw[line width=0.5pt, red, dashed] (0.309, -0.951) -- (0.1, 0.2);

\draw[line width=1.8pt] (0,0) circle[radius=1cm];

\fill[black] (0.309, 0.951) circle[radius=2pt];
\node[anchor=south west, name=n2] at (0.309, 0.951) {$e^{i\frac{2\pi}{5}}$};

\fill[black] (-0.809, 0.587) circle[radius=2pt];
\node[anchor=south east, name=n3] at (-0.809, 0.587) {$e^{i\frac{4\pi}{5}}$};

\fill[black] (0.309, -0.951) circle[radius=2pt];
\node[anchor=north west, name=n5] at (0.309, -0.951) {$e^{-i\frac{2\pi}{5}}$};

\fill[black] (-0.809, -0.587) circle[radius=2pt];
\node[anchor=north east, name=n4] at (-0.809, -0.587) {$e^{-i\frac{4\pi}{5}}$};

\fill[black] (1, 0) circle[radius=2pt];
\node[anchor=south west, name=n1] at (1, 0) {$1$};

\draw[line width=1pt, black] (1, 0) -- (0.309, 0.951);
\draw[line width=1pt, black] (0.309, 0.951) -- (-0.809, 0.587);
\draw[line width=1pt, black] (-0.809, 0.587) -- (-0.809, -0.587);
\draw[line width=1pt, black] (-0.809, -0.587) -- (0.309, -0.951);
\draw[line width=1pt, black] (0.309, -0.951) -- (1, 0);

\fill[red] (0.1, 0.2) circle[radius=2pt];

\end{tikzpicture}
\end{center}
The histogram analysis was then reduced to constructing a density scatter plot of the measurement values within the $N$-gon, with each configuration contributing a single point.

The smeared Polyakov loop presents an $N$-fold rotational symmetry, corresponding to the $\Z_N$ 1-form symmetry introduced at the beginning. This is best illustrated in 2D. Let $H$ be an $N-1$ dimensional hyperplane on the dual lattice perpendicular to the time direction. In 2D, $H$ reduces to a line. The generator of the $\Z_N$ 1-form symmetry then acts on the lattice by shifting each intersecting link by 1 in $\Z_N$. As $H$ is assumed to be orthogonal to the time direction, this implies that every temporal Polyakov loop has a link crossing $H$, and thus that $P_{\bold x} \to P_{\bold x} +1  \bmod N$ under the generator.

\vspace{0.5cm}

\begin{center}
\begin{tikzpicture}[scale=0.75]

\tikzset{
  midtriangle/.style={
    postaction={decorate},
    decoration={markings,
      mark=at position 0.45 with {
        \filldraw[blue!70] (0,0) -- (-4pt,3pt) -- (-4pt,-3pt) -- cycle;      }
    }
  }
}

\draw[gray!50] (0,0) grid (4,4);

\draw[blue!70, midtriangle] (2,0) -- (2,4);

\draw[red!70] (0,1.5) -- (4,1.5);

\node[blue!70, scale=1.4] at (2,-0.5) {$P_{\vec x}$};
\node[red!70, scale=1.4] at (-0.5,1.5) {$H$};

\draw[->] (-0.5,-0.5) to (0.5,-0.5);
\node[left] at (-0.5,0) {$t$};
\draw[->] (-0.5,-0.5) to (-0.5,0.5);
\node[below] at (0,-0.5) {$\vec x$};

\draw[->, thick] (4.5,2) -- (5.5,2)
  node[midway, above] {$\overset{s_\ell \;\to\; s_\ell+1}{}$};

\begin{scope}[xshift=6cm]
  \draw[gray!50] (0,0) grid (4,4);



  \foreach \x in {0,...,4} {
    \draw[thick,purple!90] (\x,1) -- (\x,2);
  }
\end{scope}
\end{tikzpicture}
\captionof{figure}{Illustration of the transformation 
  $s_\ell \to s_\ell + 1$ on the links crossing the hyperplane $H$. 
  The left panel shows the original configuration, and the right panel 
  highlights the links affected by the shift.}
\end{center}

\noindent This observation justifies augmenting our dataset for the smeared Polyakov histograms by including the $N$ copies of $\overline{P}$ obtained through rotations by angles of $2\pi m/ N$ with $m=0,\dots, N-1$.

\section{Reducing Floating-Point Error}
One of the big advantages of working with $\Z_N$ is that many observables can only take on a finite set of values, often of the size of $N$. This discreteness can be exploited to store measurements compactly as integer counts, avoiding cumulative sums of floating points and the rounding errors they introduce. For example, the non-connected piece in the correlator includes a term $W^*_{p_{x,\mu\nu}}W_{p_{y,\mu\nu}}$ which reduces to averages of phases of the form $e^{i\frac{2\pi ( n_x - n_y)}{N}}$. As the differences $n_x - n_y$ only take values in $\mathbb{Z}_N$, and carry all the information, storing these values encodes the data concisely and without loss of precision. We applied the same strategy to record Polyakov loops and average plaquette measurements. In the final data analysis step, the integer arrays can then be processed at the desired precision and turned into concrete measurements.

\section{Correlators}

In this section, we present a more detailed analysis of the correlators. We defined two plaquette-plaquette correlators, which we repeat here for convenience
\begin{equation}
C^\pm_{x-y;\mu\nu}=\avg{W_{p_{x,\mu\nu}}^{\pm 1}W_{p_{y,\mu\nu}}}-\big\langle W_{p_{x,\mu\nu}}^{\pm 1}\big\rangle \big\langle W_{p_{y,\mu\nu}}\big\rangle\;,
\end{equation}
where there is no sum over the repeated indices $\mu,\nu$ on the RHS. We wish to sum over indices $\mu$ and $\nu$, but here we need to distinguish between the longitudinal and transverse correlators, defining them as follows

\begin{subequations}
\begin{align}
    &C^\pm_{L}(n)=\frac{1}{4}\cdot\frac{1}{3}\sum_{\substack{\rho,\mu\\\rho\ne\mu}}C^\pm_{n\hat\rho,\rho\mu}\\
    &C^{\pm}_T(n)=\frac{1}{4}\cdot \frac{1}{3}\sum_{\substack{\rho,\mu,\nu\\\mu>\nu\ne\rho}}C^\pm_{n\hat\rho,\mu\nu}
\end{align}
\end{subequations}
where $\hat \rho$ is the unit vector in the direction of $\rho$. The factors of $\frac{1}{4}\cdot \frac{1}{3}=\frac{1}{12}$ are to account for the averages over indices. 

Note that by reflection-positivity \cite{Brydges:1979bb} we expect that $C_L^+(n)$ and $C_T^-(n)$ are each positive, so it makes sense to add them up for increased statistics. We hence define the total correlator to be
\begin{equation}
    C(n)=C_L^+(n)+C_T^-(n)\;.
\end{equation}
Up to the factor of $1/12$, this is the same correlator defined in the main article used in the ratio $C(n+1)/C(n)$.

In the main text, we presented the ratios between the correlators of subsequent lattice spacing. This analysis revealed that the fourth-order contributions dominate the long-range scaling behavior. Nevertheless, we wish to show that the curves are fitted well with the power law ansatz
\[
\sum_{k=2}^{\infty} a_{2k}\!\left(\frac{1}{x^{2k}} + \frac{1}{(L-x)^{2k}}\right),
\]

In Fig.~\ref{fig:correlators} we show the measurements of correlators $C_L^+, C_T^-$ and $C$, as well as a fitted curve up to the order $1/x^{10}$ for the total correlator. The fitting parameters are shown in Table~\ref{table:coeff}, where we skipped the 2 smallest $n$ correlator data points. 

The data is the same as the one used for Fig.~\ref{fig:power-law}, i.e. we ran Monte Carlo simulations for $\Z_3$ with $\beta=0.5114$ and $\mu =1$, for $\Z_4$ with $\beta = 1.514$ and $\tilde \beta = -0.393$, and $\Z_7$ with $\beta = 1.8$. A toric lattice of size $L=16$ was used, and we performed  $1.28\cdot 10^6$ measurements, with 500 decorrelation sweeps between them. The error was estimated using a Bootstrap.

\begin{center}
\begin{tikzpicture}
\node at (0,0) {\textbf{Raw correlators}};
\end{tikzpicture}
\begin{tikzpicture}[remember picture] 
\node[inner sep=0pt] (image3) at (0,0)
{\includegraphics[width=0.45\textwidth]{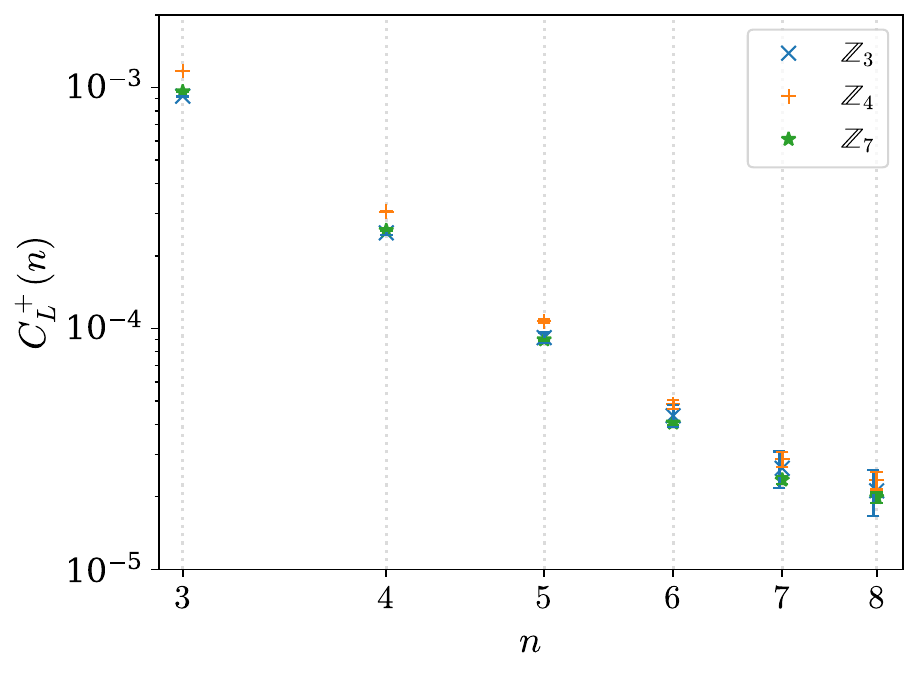}};
\end{tikzpicture}

\begin{tikzpicture}[remember picture] 
\node[inner sep=0pt] (image3) at (0,0)
{\includegraphics[width=0.45\textwidth]{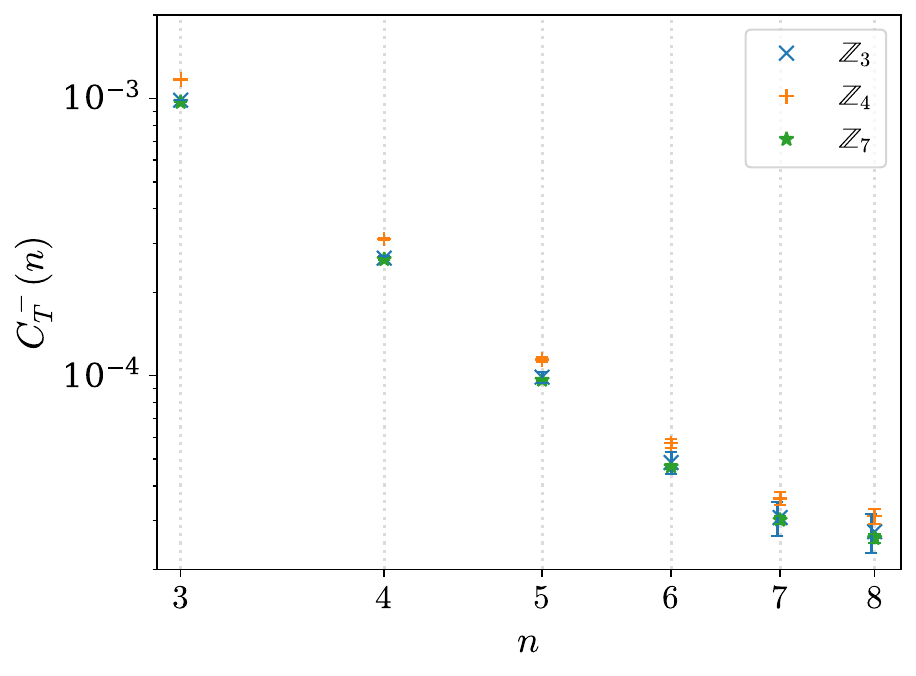}};
\end{tikzpicture}

\begin{tikzpicture}[remember picture] 
\node[inner sep=0pt] (image3) at (0,0)
{\includegraphics[width=0.45\textwidth]{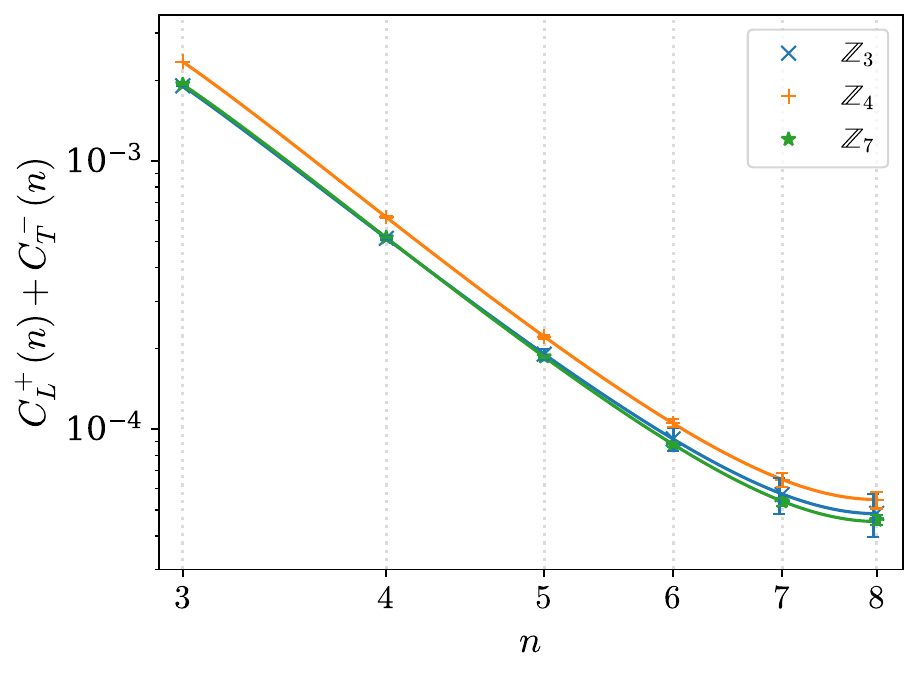}};
\end{tikzpicture}
\captionof{figure}{Raw correlators for the longitudinal, transverse, and total contributions, 
obtained from Monte Carlo simulations on a toric lattice of size $L=16$. 
Results are shown for $\mathbb{Z}_{3}$ at $\beta=0.5114$ with $\mu=1$, 
for $\mathbb{Z}_{4}$ at $\beta=1.514$ with $\tilde{\beta}=-0.393$, 
and for $\mathbb{Z}_{7}$ at $\beta=1.8$. The curve for the total correlator is obtained by fitting the series expansion up to order 10.} 
\label{fig:correlators}
\end{center}

\begin{table}[h]
\centering
\begin{tabular}{c c c c c}
\hline
$N$ & $a_{4}$ & $a_{6}$ & $a_{8}$ & $a_{10}$ \\
\hline
$\Z_{3}$ & 0.0913 & 0.4163 &  6.0466 & -42.3533 \\
$\Z_{4}$ & 0.1023 & 0.4600 & 11.1962 & -74.4772 \\
$\Z_{7}$ & 0.0825 & 0.5547 &  7.1300 & -55.5292 \\
\hline
\end{tabular}
\caption{Fitting coefficients $a_{4}$, $a_{6}$, $a_{8}$, $a_{10}$ for the total correlator.} \label{table:coeff}
\end{table}

\section{Some comments on the robustness of the Goldstone boson phase}




In this section, we contrast the Goldstone phases of $1$-form and $0$-form symmetries. The main points have been discussed in the literature (see e.g. \cite{Gaiotto:2014kfa,Hofman:2018lfz,McGreevy:2022oyu,Cherman:2023xok,Pace:2023gye}), but we review them here for completeness.

We begin by briefly reviewing the Goldstone boson phase for ordinary, 0-form symmetries. Consider a system with an \emph{exact} ordinary (i.e. $0$-form) $\U(1)$ symmetry. Further, let this system be pushed into a spontaneously broken phase, i.e. the Goldstone boson phase. The field $\varphi(x)$ corresponds to a Goldstone boson particle, i.e. an angle-valued field $\varphi\sim\varphi+2\pi$ which shifts under the exact $\U(1)$ symmetry as follows
\begin{equation}
    \U(1):\;\;\varphi(x)\rightarrow \varphi(x)+\alpha\;.
\end{equation}
The symmetry transformation parameter $\alpha$ is also angle-valued, i.e. $\alpha\sim \alpha+2\pi$, so the symmetry is a $\U(1)$ symmetry. 

Since we require the symmetry to be exact, the low-energy effective theory of the Goldstone boson must be at most derivatively coupled. Assuming full Lorentz symmetry, the effective Lagrangian is given by
\begin{equation}\label{sm:Leff_0form}
    \mathcal L_\mathit{eff} = \frac{f}{2}(\partial_\mu\varphi)^2+(\text{Higher derivative terms})\;.
\end{equation}
Since higher derivative terms are irrelevant, the effective theory is that of a free compact scalar. We should ask now: is the massless Goldstone phase stable? Firstly we need to notice that the theory actually enjoys a $(d-2)$-form symmetry following conservation of the current $j_{\mu_1 \mu_2\cdots\mu_{d-1}}=\frac{1}{2\pi}\epsilon_{\mu_1\mu_2\cdots \mu_{d-1} \mu_d}\partial^{\mu_d}\varphi(x)$. This symmetry should be considered as an emergent one, and it is explicitly broken by vortex defects. Importantly, in space-time dimension\footnote{In $d=2$ there is, of course, no strict spontaneous symmetry breaking of continuous symmetries due to the Mermin-Wagner theorem, but there still can be massless particles.} $d=2$, this is an ordinary $0$-form symmetry, and vortex operators could drive the system to a gapped phase. In higher dimensions, vortices are extended objects, and their proliferation signals a phase transition to a symmetry-restored phase. 

Let us therefore assume that the vortices are heavy or irrelevant, and that the $\U(1)$ symmetry is explicitly broken to $\mathbb Z_N$. Could a Goldstone phase emerge? To answer this question, we must check whether the Goldstone phase has relevant perturbations which reduce the symmetry down to $\mathbb Z_N$. The simplest such perturbation is by a $\cos(N\varphi)$ operator. But such deformations are relevant whenever space-time dimension $d>2$\footnote{In dimension $d=2$ the relevance of the operator depends on $f$ and its (ir)relevance is the driving mechanism of the Berezinskii-Kosterlitz-Thouless transition}, and Goldstone bosons are lifted. 

Now we repeat the argument for $\U(1)$ $1$-form symmetries. The Goldstone boson of such a symmetry is a $\U(1)$ gauge field with an effective Lagrangian
\begin{equation}\label{sm:Leff}
    \mathcal L_\mathit{eff}= \frac{1}{4e^2}\mathcal F_{\mu\nu}^2+(\text{higher gauge-invariant derivatives)}\;,
\end{equation}
where $\mathcal F_{\mu\nu}=\partial_\mu \mathcal A_\nu-\partial_\mu \mathcal A_\nu$ is the curvature tensor\footnote{Of course, $\mathcal A_\mu$ is not necessarily globally defined, and in general it must be defined in patches.}. In other words, it is just a photon (or Coulomb) phase. 

The $1$-form symmetry is a shift symmetry $\mathcal A_\mu(x)\rightarrow \mathcal A_\mu(x) +\alpha_\mu(x)$, where $\alpha_\mu(x)$ is a closed, but not exact $1$-form, i.e. a $1$-form with a vanishing curvature $\partial_\mu \alpha_\nu-\partial_\nu \alpha_\mu=0$. Note that $\alpha_\mu(x)$ is a member of the first cohomology group with coefficients in $\U(1)$, exactly what we expect for the $1$-form symmetry. Furthermore, the Wilson loops transform with a $\U(1)$ phase, i.e. $W[C]\rightarrow W[C]e^{i\int_C \alpha}$ -- where $C$ is the space-time contour of the Wilson loop -- implying that the symmetry is $\U(1)$.

Before proceeding to break the $\U(1)$ 1-form symmetry explicitly, we must ask whether the Goldstone boson theory \eqref{sm:Leff} is robust. The theory enjoys the conservation of the multi-index current $j_{\mu_1\cdots \mu_{d-2}}=\frac{1}{4\pi}\epsilon_{\mu_1\cdots\mu_{d-1}\mu_{d}}F^{\mu_{d-1}\mu_d}$, which implies that the Goldstone boson theory \eqref{sm:Leff} has an emergent $(d-3)$-form symmetry which is broken by extended $(d-3)$-dimensional defects called monopoles. Monopoles are analogous to vortices of theory \eqref{sm:Leff_0form}. The monopoles are local operators in $d=3$ and, importantly, they are relevant, gapping out the Goldstone bosons. Therefore, the remainder of the discussion will then focus on $d=4$, where monopoles are extended, particle-like objects, which can be heavy, making the Goldstone boson phase robust. 

As before, now we imagine that the $U(1)$ $1$-form symmetry is not exact, but that it emerges, enhancing the $\Z_N$ $1$-form symmetry. There is no operator we can add to \eqref{sm:Leff} to achieve this, and the simplest thing we can do is add matter fields coupled to the $\mathcal A_\mu$ gauge field. Let us add a complex scalar field $\Phi(x)$, with a covariant derivative $D_\mu\Phi=(\partial_\mu+iN\mathcal A_\mu)\Phi$, i.e. effective Lagrangian takes the form
\begin{equation}
    \mathcal L_{eff}= \frac{1}{4e^2}\mathcal F_{\mu\nu}^2+|D_\mu\Phi|^2+m^2 |\Phi|^2+\cdots
\end{equation}
Note that the choice $\alpha_\mu=\frac{\partial_\mu\xi}{N}$, with $\xi(x)\sim \xi(x)+2\pi$ an angle-valued field, is still a symmetry of the resulting effective theory if we shift $\Phi\rightarrow \Phi e^{-i\xi(x)}$. This symmetry shifts the Wilson loop $W[C]$ by a $\mathbb Z_N$ phase and is hence a $1$-form $\Z_N$ symmetry. We therefore have achieved explicit breaking of the $\U(1)$ $1$-form symmetry down to the $\Z_N$ subgroup.

Crucially, as opposed to the $0$-form symmetry Goldstone bosons, our attempt at breaking the $1$-form symmetry need not destroy the Goldstone boson phase. Indeed, if $m^2$ is large and positive, the matter field $\Phi$ can be integrated out, with minimal effect on the massless Goldstone bosons. This is why the Goldstone boson phase of the $1$-form symmetry is robust.

\section{Comment on the order of the phase transitions}

The main text discusses the existence of three generic phases of a $\Z_N$ lattice gauge theory: the spontaneously broken phase (SSB), the restored (confined) phase, and the photon phase. While the nature of the phase transitions between these phases is not the main focus of the paper, this question is potentially an important one because, if they are continuous transitions, they could potentially furnish a UV completion of an interacting $\U(1)$ gauge theory. We will, however, argue that we generically expect such transitions to be 1st order.

First, consider a transition from the SSB phase to the photon phase. Such a transition is in the same universality class as the condensation of a charge $N$ Higgs field of a $\U(1)$ gauge theory. Such a transition was studied long ago by Coleman and Weinberg \cite{Coleman:1973jx} (see also \cite{Anosova:2022yqx} for a more recent analysis) with the conclusion that it is always 1st order. 

The photon/confinement transition is also the same type of transition (this time with a charge 1 Higgs field) by the electric-magnetic duality. 

Further, in \cite{Iqbal:2021rkn} it was argued that the Ginzburg-Landau theory of 1-form symmetries generically has a cubic term, and the authors believe this causes all transitions involving a 1-form symmetry to generically be 1st order. Further still, if the transition is 2nd (or higher) order, it must flow to a conformal fixed point in the IR. However, in \cite{Hofman:2018lfz} it was argued that a conformal field theory with a $\U(1)$ $1$-form symmetry (expected to emerge at the fixed point) must flow to a non-interacting photon phase, which has no relevant deformation driving it to a gapped phase. 

A potential loophole is a transition where both electric and magnetic matter is light. This was the question asked in ref. \cite{Anosova:2022yqx}, but no such fixed point was found with a single species of electric and magnetic matter. Nevertheless, the question remains an open one.

By the above arguments on fairly general grounds, we expect that all transitions observed in the $\Z_N$ lattice gauge theory are always 1st order, although we cannot preclude more exotic scenarios. The most interesting case to study would be at the point where all three phases meet, and where a strongly interacting fixed point cannot be a priori excluded.

\end{appendix}

\end{document}